\documentclass[showpacs,aps,prb,twocolumn,superscriptaddress]{revtex4-1}

\usepackage{graphicx}  
\usepackage{subfigure}
\usepackage{dcolumn}   
\usepackage{bm}        
\usepackage{amssymb}   
\usepackage{comment}
\usepackage{hyperref}
\usepackage{color}
\usepackage{amsmath}
\usepackage{mathrsfs}
\usepackage{mathcomp}
\usepackage{textcomp}
\usepackage{dsfont}
\usepackage{esint}
\usepackage{braket}
\usepackage{textgreek}
\usepackage{lipsum}

\begin{document}

\title{Instability and topological robustness of Weyl semimetals against Coulomb interaction}

\author{Fei Xue}
\email{feixue@utexas.edu}
\affiliation{Department of Physics, The University of Texas at Austin, Austin, TX 78712, USA}
\author{Xiao-Xiao Zhang}
\email{zhang@appi.t.u-tokyo.ac.jp}
\affiliation{Department of Applied Physics, The University of Tokyo, 7-3-1 Hongo, Bunkyo-ku, Tokyo 113-8656, Japan}


\newcommand\dd{\mathrm{d}}
\newcommand\ii{\mathrm{i}}
\newcommand\ee{\mathrm{e}}
\newcommand\zz{\mathtt{z}}
\makeatletter
\let\newtitle\@title
\let\newauthor\@author
\def\ExtendSymbol#1#2#3#4#5{\ext@arrow 0099{\arrowfill@#1#2#3}{#4}{#5}}
\newcommand\LongEqual[2][]{\ExtendSymbol{=}{=}{=}{#1}{#2}}
\newcommand\LongArrow[2][]{\ExtendSymbol{-}{-}{\rightarrow}{#1}{#2}}
\newcommand{\cev}[1]{\reflectbox{\ensuremath{\vec{\reflectbox{\ensuremath{#1}}}}}}
\newcommand{\red}[1]{\textcolor{red}{#1}} 
\newcommand{\mycomment}[1]{} 
\makeatother

\begin{abstract}
There is a close 
connection between various new phenomena in Weyl semimetals and the existence of 
linear band crossings in the single particle description. We show, by a full self-consistent mean-field calculation, how this picture is modified in the presence of long-range Coulomb interactions. The chiral symmetry breaking occurs at strong enough interactions and 
the internode interband excitonic pairing channel is found to be significant, which determines the gap-opened band profile varying with interaction strength. Remarkably, in the resultant interacting phase, finite band Chern number jumps in the three-dimensional 
momentum space are retained, indicating the robustness of the topologically nontrivial features.
\end{abstract}

\maketitle



\section{Introduction}%
The physics related to the Weyl point or Weyl node, as the three-dimensional (3D) analog of the two-dimensional (2D) Dirac physics\cite{DiracFermion1,*DiracFermion2}, is sparking keen interests both theoretically and experimentally\cite{ReviewBurkov,WeylDiracReview}. This started from the revival of the old concept of Weyl fermion\cite{Weyl1929} in the context of various condensed matter systems without time-reversal symmetry and/or inversion symmetry\cite{Volovik,Weyl2007,Weyl2011}. Besides the recent solid-state realizations in a family of nonmagnetic and noncentrosymmetric transition metal monoarsenides/monophosphides 
\cite{predict2,*predict1,*TaAS1,*TaAS2}, it is found or predicted as well in photonic crystals\cite{WeylPhotonic1}, magnon bands\cite{WeylMagnon1,*WeylMagnon2,*WeylMagnon3,*WeylMagnon4}, and even photo-driven systems\cite{\mycomment{Xing,*}XXZ1,*ZhongWang,*Rubio\mycomment{,*DanielLoss,*Awadhesh,*Oka}}. In contrast to the real-space emergent monopole structures\cite{spinice0,*spinice1,*XXZ:resistivity,*XXZ:monopole}, a Weyl node has a momentum-space monopole nature\cite{Volovik,EEMF0}. Based on this and its special Landau level formation under a magnetic field\cite{Nielson-Ninomiya1,*Nielson-Ninomiya2}, many new phenomena are discussed and experimentally investigated in this gapless quantum phase of matter, i.e., the Weyl semimetal, including but not limited to the anomalous Hall effect\cite{AHE1,AHE2}, the chiral magnetic effect\cite{CME0,CME1,CME2}, and the observation of the negative magnetoresistance\cite{seeCMEDirac1,*seeCMEDirac2,seeCMEWeyl1,*seeCMEWeyl2}. 

The Weyl nodes as degeneracies of codimension three are generally stable against local perturbations unless opposite-chirality nodes separated in the momentum space are 
coupled to break the chiral symmetry. The disorder might not open a gap either, due to the randomly vanishing pinned Fourier component and the inadequate strength in practice. The other indispensable aspect is the interaction effect. There are some theoretical studies with different focuses and approximations to facilitate analytic analyses, including mean-field or renormalization group calculations within or beyond local interactions\cite{WeylCDW1,Nandkishore,Sekine,Aji2014,Juricic2017} and 
the formation of spin or charge density waves\cite{WeylSDW,Ran_model,WeylCDW2,WeylCDW3} or Luttinger liquids\cite{Meng,XXZ2} under an external magnetic field.
On the other hand, it is well known that an excitonic semimetal-insulator phase transition could occur under the influence of strong enough long-range Coulomb interaction\cite{Keldysh1965,Kohn1967,Lozovik1976,Khveshchenko2001,cutoff2}. The Coulomb interaction can bind electrons and holes to excitons, quasibosonic bound states, which can condense at low temperatures. Previous exciton condensate studies focus on semiconductor bilayer systems \cite{Lozovik1976,Zhu1995,FX1} and a 2D quantum well placed in an optical cavity\cite{Deng2002, Kasprzak2006,Snoke2007,Keeling2007,FX2} because a Bose-Einstein condensate (BEC) type low density exciton (polariton) limit exists in these 2D systems. Can we apply some similar analysis to a Weyl semimetal and address the natural question of whether it 
is stable against the Coulomb interaction and in what sense?

To answer this, we study a simple type-I Weyl semimetal with vanishing density of states, i.e., the chemical potential is tuned at the Weyl points, under the long-range Coulomb interaction upon which we apply a standard Hartree-Fock approximation. The theory would become less valid when the chemical potential is away from the Weyl points due to complex interaction effects beyond the mean-field level, such as the strong screening in 3D at a finite density of states which renormalizes the interaction down to a short-range form\cite{Lindhard1954, NozieresPines1999, GiulianiVignale2008, GrapheneRMP2012}.  Within the mean-field level taking all possible electron-hole pairing channels into account, we numerically carry out the self-consistent calculation without any other \textit{a priori} approximation in order to draw unambiguous conclusions.

Our main findings are twofold. First, a strong enough long-range Coulomb interaction connects the left and right nodes and breaks the chiral symmetry and the translational symmetry as well, leading to a finite gap opening. We explain how the quasiparticle band profile evolves using two 
order parameters, the dressed single particle band energy and the internode interband coupling. Second, using the self-consistent Hamiltonian obtained to calculate Chern numbers of many 2D slices in the 
momentum space, we find that the nontrivial topology of the system is robust and retained despite losing the Weyl nodes.
This supports the topologically nontrivial and axionic nature of the density wave phase\cite{thetaWeyl,WeylCDW2,WeylCDW3,Nandkishore}.

The paper is organized as follows. In Sec.~\ref{Sec:Model} we introduce our 
model Hamiltonian 
and explain how we apply mean-field calculations
. In Sec.~\ref{Sec:Results} we present our band structure results and Chern number calculations where the phase transition occurs under strong enough interactions. We also discuss the quasiparticle band evolution and demonstrate that the transition is continuous. In Sec.~\ref{Sec:Conclusion} we summarize  and comment on related issues.



\section{Model}\label{Sec:Model}
We consider a general continuum model of a time-reversal symmetry breaking Weyl semimetal with two Weyl nodes located symmetrically at $\pm\vec{K}=\pm K \hat{z}$ ($\hat{z}$ is defined in momentum space rather than real space), labeled as $s=R/L$ node or interchangeably $s=\pm$, respectively. 
The Hamiltonian can be written in the basis spanned by the state $\ket{\beta_\sigma^s}$ of the Weyl electron (pseudo)spin $\sigma=\uparrow/\downarrow$ and the node pseudospin $s=R/L$ 
\begin{equation}\label{eq:Hamiltonian}
\hat{H}_0=\sum_{\vec{k}}\psi_{\vec{k}}^{\dagger}
\hbar v_F \vec{k}\cdot \vec{\sigma} s^z
\mycomment{\begin{pmatrix}
\hbar v \vec{k}\cdot \vec{\sigma} & 0\\
0 & -\hbar v_F \vec{k}\cdot \vec{\sigma}\\
\end{pmatrix}}
\psi_{\vec{k}}
+\hat{H}_\mathrm{I},
\end{equation}
where $\vec{k}$ is the reduced momentum relative to the nodes at $\pm K \hat{z}$ and $v_F$ is the Fermi velocity which varies in different materials, $\vec{\sigma}$ is the electron (pseudo)spin Pauli matrix, $\vec{s}$ is the node pseudospin
Pauli matrix, $\psi_{\vec{k}}=(c_\uparrow^R,c_\downarrow^R,c_\uparrow^L,c_\downarrow^L)^T$, and
the electron annihilation/creation operator $c_\sigma^{s(\dag)}$  is for the state $\ket{\beta_\sigma^s}$ with $\ket{\beta_\uparrow^s}=(1,0)^T\,,\ket{\beta_\downarrow^s}=(0,1)^T$.

Taking account of the presence of Weyl nodes (Appendix~\ref{App:MeanField}), the Coulomb interaction for this system has the form
\begin{equation}\label{eq:interaction_Hamiltonian}
\begin{split}
\hat{H}_\mathrm{I}=\frac{1}{2\Omega}\sum_{\sigma\sigma',ss'}\sum_{\vec{k},\vec{k}',\vec{q}}
[V(\vec{q}) c_{\sigma \vec{k}+\vec{q}}^{s\dagger} c_{\sigma' \vec{k}'-\vec{q}}^{s'\dagger} c_{\sigma' \vec{k}'}^{s'} c_{\sigma \vec{k}}^{s} \\
+ V(\vec{q}+2s\vec{K}) s_{ss'}^x c_{\sigma \vec{k}+\vec{q}}^{s\dagger} c_{\sigma' \vec{k}'-\vec{q}}^{s'\dagger} c_{\sigma' \vec{k}'}^{\bar{s'}} c_{\sigma \vec{k}}^{\bar{s}}] 
\end{split}
\end{equation} 
where $\Omega$ is the 3D system volume, 
$V(\vec{p})=e^2/(\varepsilon_0\varepsilon_r |\vec{p}|^2)$ with the vacuum/relative permittivity $\varepsilon_0 / \varepsilon_r$, and $\bar{s}\neq s$. 
Equation~\eqref{eq:interaction_Hamiltonian} includes all possible intranode/internode scattering processes allowed by momentum conservation.
This reduced momentum representation has the merit of expressing all the mean-field interactions in a momentum-diagonal manner.
Applying the Hartree-Fock approximation\cite{Fetter,Mahan} to the interaction Hamiltonian in Eq.~\eqref{eq:interaction_Hamiltonian} as detailed in Appendix~\ref{App:MeanField}, we finally get
\begin{equation}\label{eq:MF}
\hat{H}_\mathrm{MF}=\sum_{\vec{k}} \psi_{\vec{k}}^{\dagger}
(H_0+H_\mathrm{Hartree}+H_\mathrm{Fock})
\psi_{\vec{k}}
\end{equation}
where 
\begin{equation}\label{eq:Hatree}
\left[H_\mathrm{Hartree}\right]_{\sigma\sigma'}^{ss'} = V(2\vec{K})\sum_{\vec{k}'}(\rho^{ss'}_{\sigma\sigma'}+\rho^{ss'}_{\bar{\sigma}\bar{\sigma}'})_{\vec{k}'} \delta_{\sigma\sigma'} s^x_{ss'},
\end{equation}
and
\begin{equation}\label{eq:exchagne}
\begin{split}
\left[H_{\mathrm{Fock}}\right]_{\sigma\sigma'}^{ss'} = -\sum_{\vec{k}'}[V(\vec{k}-\vec{k}')\rho^{ss'}_{\sigma\sigma'\vec{k}'} \\
+ \delta_{ss'} V(\vec{k}-\vec{k}' + 2s\vec{K})\rho^{\bar{s}\bar{s}'}_{\sigma\sigma'\vec{k}'}].
\end{split}
\end{equation}
Here, $\bar{\sigma}\neq \sigma$, and the density matrix is defined relative to a reference value determined by the filling of the noninteracting ground state
$\rho^{ss'}_{\sigma\sigma'\vec{k}} = \braket{c_{\sigma' \vec{k}}^{s'\dagger} c_{\sigma \vec{k}}^s}
-\bar{\rho}^{ss'}_{\sigma\sigma'\vec{k}}$,
where $\bar{\rho}^{ss'}_{\sigma\sigma'\vec{k}} = \delta_{ss'}\braket{\beta_v^s|\beta_{\sigma'}^s} \braket{\beta_{\sigma}^s|\beta_v^s}$ (see the next paragraph). 
We note that the long-range Coulomb interaction allows for both the internode and intranode couplings. Even though the Hartree contribution to the internode coupling might be small due to the decay of the Coulomb interaction at large momentum transfer, the Fock contribution still accommodates possible strong internode coupling. In our self-consistent calculation, given all nonzero initial terms, the density matrix $\rho$ and hence the mean-field Hamiltonian $\hat{H}_\mathrm{MF}$ are iteratively updated at each iteration using the lowest two eigenvectors until the convergence. Note that the trace of $\rho$ at each momentum is always zero.

For each node, the noninteracting system can be diagonalized in the band basis representation $\ket{\beta_n^s}$ of band $n=c,v$ (conduction/valence band) and node $s$. We have $\ket{\beta_c^R}_{\vec{k}}=\ket{\beta_v^L}_{\vec{k}}=(\cos{\frac{\theta}{2}},\sin{\frac{\theta}{2}}\ee^{\ii\phi})^T$ and $\ket{\beta_v^R}_{\vec{k}}=\ket{\beta_c^L}_{\vec{k}}=(-\sin{\frac{\theta}{2}},\cos{\frac{\theta}{2}}\ee^{\ii\phi})^T$, where the momentum $\vec{k}$ has polar and azimuth angle $\theta,\phi$. 
In order to avoid the tedious band states' overlap functions, as shown above, we formulate the interaction Hamiltonian in the spin basis $\ket{\beta_\sigma^s}$ where we take care of the noninteracting ground state using
the relative density matrix $\rho$.
Since we include all the possible interaction channels, it is fundamentally equivalent to work with either the spin basis or the band basis, which are related by a unitary transformation $U^{ss'}_{n\sigma}=s^0_{ss'}\braket{\beta_n^s|\beta_\sigma^{s'}}$.
For instance, we actually determine the aforementioned reference matrix $\bar{\rho}$ and the form of $H_\mathrm{Hartree}$ using $U$ and the overall charge neutrality constraint $\sum_{n} \braket{c_{n \vec{k}}^{s\dagger} c_{n \vec{k}}^s} = 1$.
Especially, we will later use the band basis density matrix and Hamiltonian (henceforth denoted with a tilde) of the form $\tilde{\rho} = U^{\dag T} \rho U^T$ and $\tilde{H} = U H_\mathrm{MF} U^\dag$, whereupon more transparent physical understandings become available.

We use a modified Rydberg unit in the calculation, setting $\hbar=\frac{e^2}{2}=4\pi\varepsilon_0=1$ where $\hbar$ is the Planck constant, $e$ is the elementary charge, and $\varepsilon_0$ is the vacuum permittivity. This leads to a characteristic velocity $v_0=\frac{4\pi\varepsilon_0 e^2}{2\hbar}=1$ ($10.9\times 10^5 m/s$ in SI unit). Combined with the material dependent relative permittivity $\varepsilon_r$ and the Fermi velocity $v_F$, we have the quantity $\frac{v_0}{\varepsilon_rv_F}$ characterizing the strength of the interaction. Therefore, smaller $v_F$ and $\varepsilon_r$ effectively mean stronger Coulomb interaction effects and only the combined value of $\varepsilon_r v_F$, referred to as the relative velocity $v_r$ henceforth, matters. The realistic typical ranges are $v_F=0.5\text{--}3\times 10^5 m/s$ and $\varepsilon_r=10\text{--}20$\cite{predict3,*predict4,dielectric1,*dielectric2}, which means about $v_r=0.5\text{--}6$ in our unit.
Because of the massless linearly dispersed band structure and the long range Coulomb interaction, an important feature of this system is the lack of an intrinsic length 
scale\cite{cutoff1,cutoff2,GrapheneRMP2012}. Even if one sums up to a certain momentum magnitude (a sharp ultraviolet bandwidth cutoff $v_rK$ in our case), the obtained band energies will be just proportional to that cutoff. Therefore, in our theory, the concrete predictions are the band profiles, phase transitions, topological features rather than the exact gap or band energies. Indeed, we observe this feature in our numerical calculations and an inspection of its self-consistency is given in Appendix~\ref{App:SelfCon}.

\section{Results}\label{Sec:Results}
Our numerical results are based on a $34\times34\times34$ cubic momentum grid with $k_x,k_y,k_z \in [-K/2,K/2]$. 
The momenta and energies (since $v_0=1$) are thus indicated with a unit $k_c=K/2$ in all the figures.
The minimal momentum magnitude is therefore $k_\Gamma=K/66$ and up to this accuracy, we refer to momenta along an axis or at the $\Gamma$ point in the following. Here we deliberately detour the $\Gamma$ point to avoid the gauge choice ambiguity at the node. In this setup, left and right nodes located at $\pm K \hat{z}$ are well separated and expressed in the diagonal blocks in the $4\times4$ Hamiltonian. Only the long-range Coulomb interaction can induce off-diagonal terms that lead to chiral symmetry breaking and gap opening. Tuning the relative velocity $v_r$ as aforementioned and keeping Coulomb interaction in its vacuum form, we can explore different phases and physical properties by looking at the renormalized quasiparticle band structure and the Chern number profile.

\subsection{Mean-field band profile}\label{Sec:Band}

\begin{figure}[t]
	\includegraphics[width=1\columnwidth]{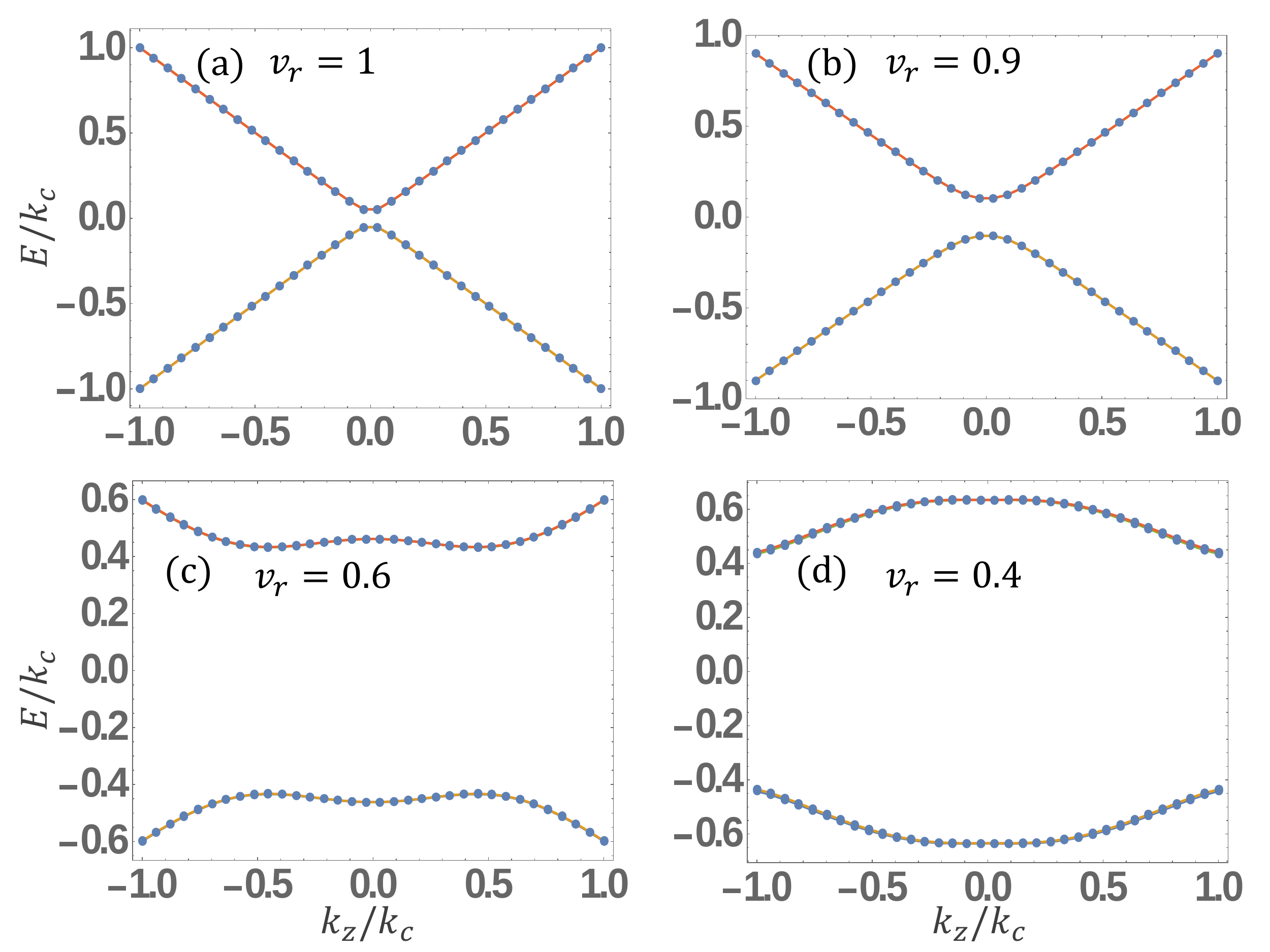}
	\caption{(Color online) Quasiparticle energy bands along $k_z$ at four relative velocities $v_r=1.0, 0.9, 0.6, 0.4$.
		The blue dots denote four eigenvalues of our self-consistently converged mean-field Hamiltonian.
		Color lines are guides to the band profile. Double degeneracy is lifted although too small to be visible in (b) to (d).
	}
	\label{Fig:band}
\end{figure}

\begin{figure}[t]
	\includegraphics[width=1\columnwidth]{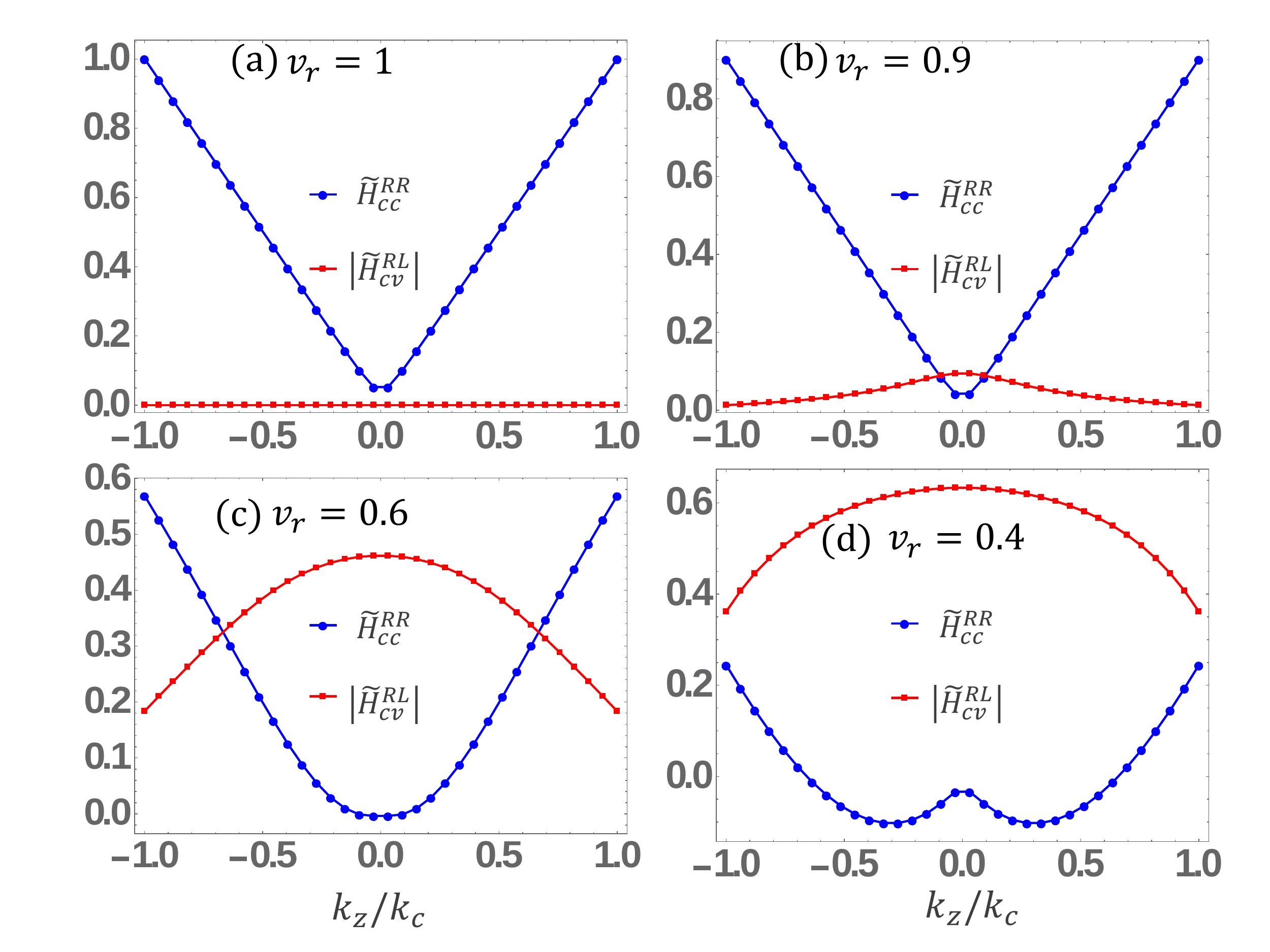}
	\caption{(Color online) Interband internode coupling magnitude $|\tilde{H}_{cv}^{RL}|$ and one diagonal term $\tilde{H}_{cc}^{RR}$ along the $k_z$ axis at relative velocities $v_r=1.0, 0.9, 0.6, 0.4$.
		The blue dots  and red squares denote $\tilde{H}_{cc}^{RR}$ and $|\tilde{H}_{cv}^{RL}|$, respectively.}
		\label{Fig:H14}
	\end{figure}
In Fig.~\ref{Fig:band}, we plot renormalized four eigenvalues along 
the $k_z$ axis. The plots along $k_x$ and $k_y$ axes are the same for the unbroken rotational symmetry with respect to the $k_z$ axis, and their difference with Fig.~\ref{Fig:band} is insignificant. Unlike in the 2D case where arbitrarily weak attractive interaction will create bound states, a strong enough interaction is required to create internode or intranode electron-hole bound states, if any, in 3D Weyl semimetals. At a large relative velocity $v_r=1$, the interaction strength is not strong enough to bind electron-hole pairs and the band profile is unchanged, i.e., $E=\pm|\vec{k}|$ with double degeneracy. Note that the tiny gap in Fig.~\ref{Fig:band}(a) simply comes from the momentum resolution $k_\Gamma$ and all the energies have perfect linearity. To clarify whether we have electron-hole excitonic pairs, we also plot the dominant interaction induced internode electron-hole pairing term magnitude $|\tilde{H}_{cv}^{RL}|$  
along $k_z$ shown as red dots in Fig.~\ref{Fig:H14}. Note that we have $\tilde{H}_{cv}^{RL}=\tilde{H}_{cv}^{LR}$ due to the inversion symmetry. 
At $v_r=1$, $\tilde{H}_{cv}^{RL}$ in Fig.~\ref{Fig:H14}(a) converges to zero everywhere, illustrating the absence of any bound states.

Other than this case, we observe gap openings in Fig.~\ref{Fig:band} which are accompanied by the strong internode s-wave pairings, which are nonzero at zero momentum, shown in Fig.~\ref{Fig:H14}. 
Meanwhile, a finite number of electron-hole pairs are created by the interaction and 
the exciton density reads $n_{ex}=\mathrm{tr}\tilde{\rho}_{cc}/\Omega=\sum_{\vec{k}}(\tilde{\rho}_{cc\vec{k}}^{RR}+\tilde{\rho}_{cc\vec{k}}^{LL})/\Omega$, which equals the similar valence band trace of $\tilde{\rho}_{vv}$ by virtue of the particle-hole symmetry. Because we place the Weyl nodes on the $k_z$ axis, interaction terms containing $2K \hat{z}$ self-consistently persist. It not only breaks the rotational symmetry with respect to the $k_x$ or $k_y$ axes, but also, more importantly, completely lifts the double degeneracy, although the splitting is small due to the decay of the Coulomb potential at large momenta. For instance, we have $\tilde{\rho}_{cc}^{RR} \neq \tilde{\rho}_{cc}^{LL}$ in consequence.

\begin{figure}[t]
	\includegraphics[width=1\columnwidth]{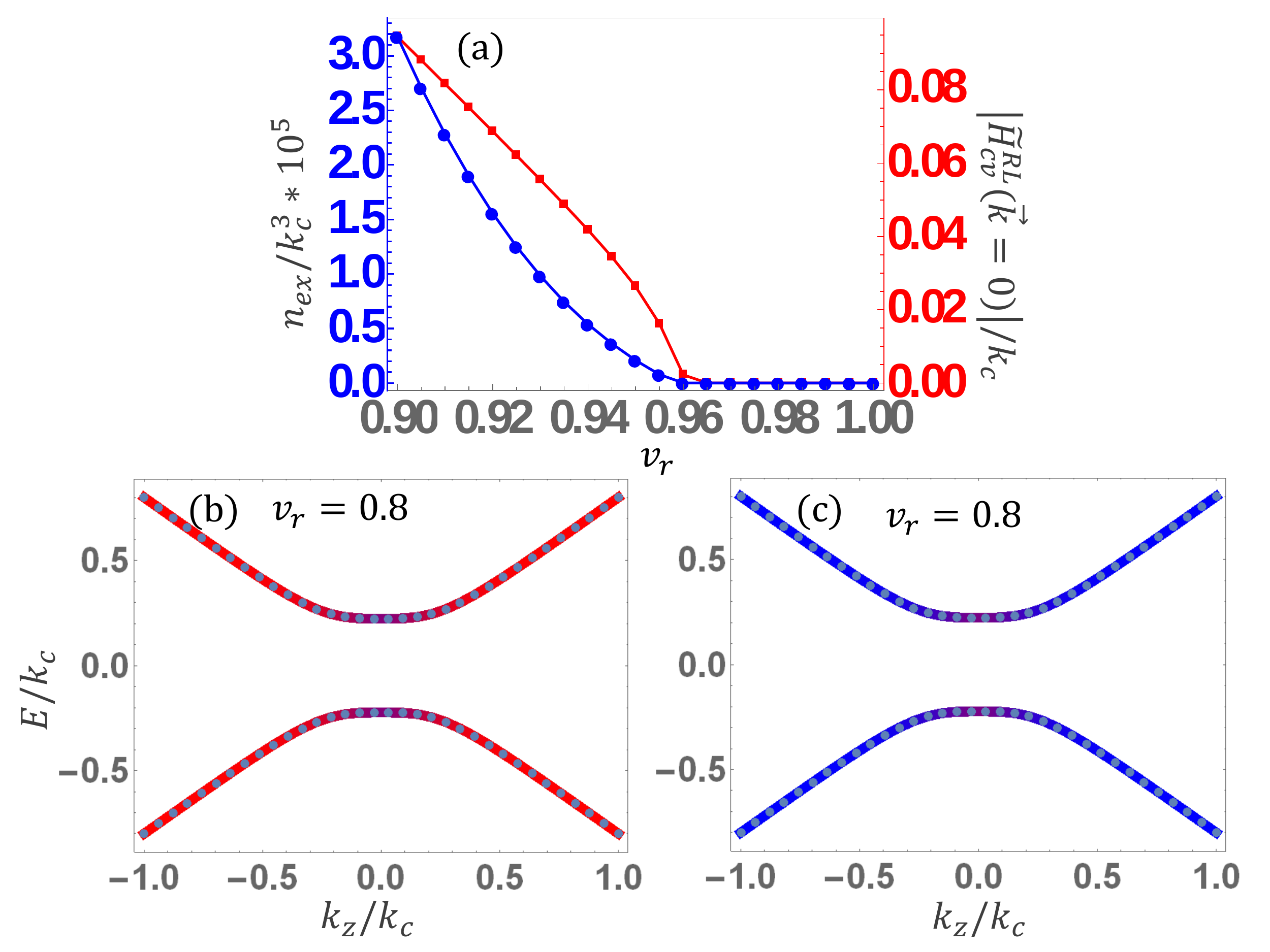}
	\caption{(Color online) (a) Order parameter $\tilde{H}_{cv}^{RL}(\vec{k}=0)$ (red dots) and exciton density $n_{ex}=\mathrm{tr}\tilde{\rho}_{cc}/\Omega$ (blue squares) as a function of the relative velocity $v_r$.
	$\tilde{H}_{cv}^{RL}(\vec{k}=0)$ is averaged using the eight points $(\pm k_{\Gamma},\pm k_{\Gamma},\pm k_{\Gamma})$ nearest to the $\Gamma$ point, and $n_{ex}$ is magnified $10^5$ times for better
	illustration.
	Node weight of (b) the second and third highest bands and (c) the first and fourth highest bands along the $k_z$ axis after the phase transition. Blue (red) represents weight from the left (right) node.
	}
	\label{Fig:phasetransition}
\end{figure}

To clearly inspect this phase transition, in Fig.~\ref{Fig:phasetransition}(a), we plot two order parameters, the internode interband pairing magnitude $|\tilde{H}_{cv}^{RL}(\vec{k}=0)|$ and the exciton density $n_{ex}$ in a small range of $v_r$ near the critical value $v_r^c=0.96$, which is similar to a recent study\cite{Zong2017}. 
Both $|\tilde{H}_{cv}^{RL}(\vec{k}=0)|$ and $n_{ex}$ show a typical second-order continuous 
phase transition that is expected from normal exciton condensates\cite{Keldysh1965,Lozovik1976,Comte1982}.
Using the wavefunctions, we also plot bands with node weight in Figs.~\ref{Fig:phasetransition}(b) and \ref{Fig:phasetransition}(c) to illustrate the lifting of the R/L node degeneracy and the no longer conserved node pseudospin under strong interaction. One particle-hole symmetric pair of bands (second and third highest bands) chooses one major node while the other (first and fourth highest bands) chooses the opposite. At small momenta, R/L nodes are mixed in agreement with later discussion that the interaction induced internode pairing mainly contributes at small momenta when the phase transition just occurs. When interaction is not strong enough, these pairs of bands have entirely mixed colors in calculation due to the R/L degeneracy.

When we further decrease $v_r$, the quasiparticle gap becomes larger and the band extrema move from zero to finite momenta relative to the cutoff momentum. This hump-like profile is similar to the BEC-BCS crossover\cite{Keldysh1965,Lozovik1976,Comte1982} where exciton condensate pair-excitation energy extrema move to finite momenta at larger exciton density. To illustrate band evolution after the phase transition, we write down a simplified Hamiltonian in the band basis based on our mean-field result and the symmetry,
\begin{equation}
\label{eq:effectiveH}
\tilde{H}'=
\begin{pmatrix}
v_rk-\xi_{\vec{k}} &0&0 & \Delta_{\vec{k}}\\
0 &-v_rk+\xi_{\vec{k}}&\Delta^*_{\vec{k}} & 0\\
0 &\Delta_{\vec{k}}&v_rk-\xi_{\vec{k}}  & 0\\
\Delta^*_{\vec{k}} &0&0 & -v_rk+\xi_{\vec{k}}
\end{pmatrix},
\end{equation}
where we only consider the internode interband coupling $\Delta_{\vec{k}}$ and the dressed single particle energy $\xi_{\vec{k}}$, i.e., $\tilde{H}_{cv}^{RL}$ and $\tilde{H}_{cc}^{RR}$ previously discussed, and neglect the degeneracy-lifting effect. Both the intranode interband and internode intraband terms are negligibly small and dropped for simplicity. 
The corresponding eigenvalues are doubly degenerate $E_{\vec{k}}=\pm\sqrt{(v_rk-\xi_{\vec{k}})^2+|\Delta_{\vec{k}}|^2}$. The pair-excitation gap at each momentum is $2|E_{\vec{k}}|$.
At $v_r=0.9$ in Fig.~\ref{Fig:band}(b) where the phase transition just occurs, a small amount of internode electron-hole pairs are bound and exciton condensates form. The condensates lead to the small but finite internode interband pairing and the dressing of single-particle energy. 
As shown in Fig.~\ref{Fig:H14}(b), $\Delta_{\vec{k}}$ is maximized at $k=0$ and decays with increasing $k$. Since the single particle part $v_r k$ now outweighs $\xi_{\vec{k}}$, $v_r k - \xi_{\vec{k}}$ remains almost linearly increasing with slope $v_r$. 
The gap minimum is thus located at zero momentum as a result of the dominant single-particle contribution.
At large momenta, since the interaction effect is diminished, the band structure resembles the noninteracting linear one.

As $v_r$ becomes even smaller shown in Fig.~\ref{Fig:H14}(c), $\Delta_{\vec{k}}$ outweighs the diagonal part $v_r k-\xi_{\vec{k}}$ at small momenta due to the cancellation between $v_r k$ and $\xi_{\vec{k}}$. 
The gap minimum then moves to finite momenta because $\Delta_{\vec{k}}$ decreases slowly as the momentum increases. At large momenta, the noninteracting term again becomes more pronounced, leading to hump like bands in Fig.~\ref{Fig:band}(c). 
For completeness, we also plot the case of very small $v_r$ in Fig.~\ref{Fig:band}(d) in which the top (bottom) of the upper (lower) band is at small momenta. 
Unlike the previous ones, it seemingly inherits little linear band remnants.
This originates from the fact that the single particle linear band is not strong enough to diminish the Coulomb interaction to small values at large momenta, which is seen from the dominance and slow decay of the interaction induced $\tilde{H}_{cv}^{RL}$ in Fig.~\ref{Fig:H14}(d). 
Then $\Delta_{\vec{k}}$ and hence the gap decrease as the momentum increases. 
If one were to avoid this, nonlinearity could be introduced to the Weyl bands at large momenta, which is, however, not within the scope of the current study. 
Last but not least, we emphasize the importance of including $\xi_{\vec{k}}$, which is from the intraband exchange interaction, in a full mean-field calculation in order to explain how renormalized quasiparticle bands evolve. In some previous studies, it was ignored to analyze the pairing gap alone\cite{Aji2014,Wang2017}, which is unjustified in a complete theory and insufficient to capture all the physics such as hump like quasiparticle bands.

\subsection{Topological property}\label{Sec:Topo}
\begin{figure}[t]
	\includegraphics[width=1\columnwidth]{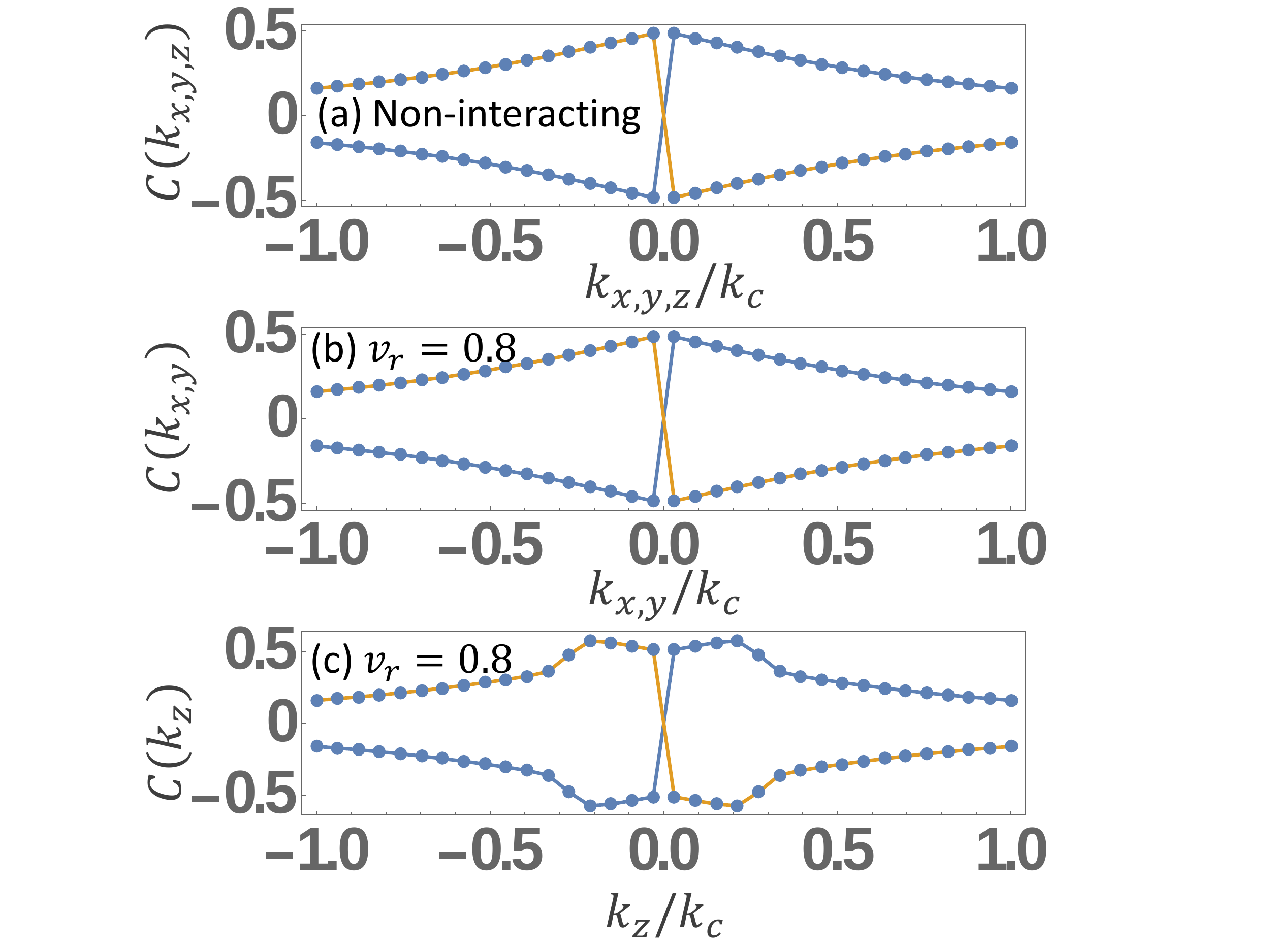}
	\caption{(Color online) Chern number $C$ for the lowest two bands for (a) the noninteracting case as a function of $k_{x,y,z}$ and the interacting case when $v_r=0.8$ as a function of (b) $k_{x,y}$ and (c) $k_z$. For the sake of comparison, we calculate the exactly soluble case (a) under the same condition of momentum cutoff and sampling mesh as (b) and (c). 
	}
	\label{Fig:Chern}
\end{figure}

Many of the topologically nontrivial features of the Weyl semimetal stem from the Weyl node as a source or sink of the flux of the momentum-space Berry phase. The simple and thorough way to see is to scan the Brillouin zone and calculate the Chern numbers slice by slice along several directions, say, $k_x,k_y,k_z$. A Chern number jump along any direction from $\pm\frac{1}{2}$ to $\mp\frac{1}{2}$ clearly indicates the presence of a Weyl node of charge $\mp1$. Here we will answer the question over the fate of such topological properties in the presence of the interactions. For any continuum model, the exact quantization will never be attainable unless one pushes the range of momentum toward the infinity. Taking the $k_z$-slice ($k_x$-$k_y$-plane) Chern number $C(k_z)$ of the noninteracting Weyl semimetal 
as an example, practically despite the imperfect quantization and the decay for larger and larger $|k_z|$ as shown in Fig.~\ref{Fig:Chern}(a), one can still identify a sharp jump of about $\pm1$ at the node position.

However, since one typically has to sum up the Berry curvature over the momentum space using the TKNN-type Kubo formula\cite{ChernTKNN1,*ChernTKNN2}, there lies another severe problem, viz., the density of the sampling mesh, which is in general very limited in 3D numerical calculations. One way out is the Wilson loop method that counts the winding of the Wannier center in a cylinder geometry and applies to various distinct topological systems\cite{ChernFang1,*ChernFang2}. Here, to make the most direct use of our calculated data, we adopt another strategy of remarkably fast convergence even with a 2D momentum mesh of tens or hundreds of points\cite{ChernHatsugai}. Based on lattice gauge theory, it sums up the gauge invariant plaquette Berry flux, e.g., for the $k_x$-$k_y$ plane Chern number, 
\begin{equation}
\mycomment{C = \frac{1}{2\pi}\Im\sum_{\vec{k}}{\ln[U_x(\vec{k}) U_y(\vec{k}+\hat{x}) U_x(\vec{k}+\hat{y})^{-1} U_y(\vec{k})^{-1}]},}
C = \frac{1}{2\pi}\Im\sum_{\vec{k}}{[A_x(\vec{k}) + A_y(\vec{k}+\hat{x}) -A_x(\vec{k}+\hat{y}) - A_y(\vec{k})]},
\end{equation}
where the summation is over a discrete momentum mesh and the lattice Berry connection $A_i(\vec{k})=\ln[\braket{\Psi(\vec{k})|\Psi(\vec{k}+\hat{i})}/\lvert\braket{\Psi(\vec{k})|\Psi(\vec{k}+\hat{i})}\rvert]$ with the normalized Bloch state $\ket{\Psi(\vec{k})}$ solved from our self-consistently converged Hamiltonian.

Surprisingly, as shown in Figs.~\ref{Fig:Chern}(b) and \ref{Fig:Chern}(c), up to a $3\%$ deviation from $\pm1$, we find that each of the four bands retain the Chern number jump along every direction at the original node position, i.e, the $\Gamma$ point of the reduced momentum. This appears identically in cases after the phase transition whereas we show only the $v_r=0.8$ one for simplicity. $C(k_{x,y})$, unlike $C(k_z)$, has the same shape as the noninteracting case, which is again due to the asymmetry between $k_z$ and $k_{x,y}$. Also, adjacent bands possess opposite slice Chern number at any momentum and hence opposite jumps. This means that, despite the fact that the R/L node is no longer a good quantum number due to the interaction induced mixing, the bands still partially inherit the topological features. 
As suggested by some previous studies, the axionic character and hence the $\theta$ term due to the chiral anomaly can survive from the dynamical mass generation due to the chiral symmetry breaking\cite{thetaWeyl,WeylCDW2,WeylCDW3,Nandkishore}. Here we demonstrate directly from a topological number calculation that the band topology is indeed more robust than the gapless Weyl nodes themselves. Thus, the Coulomb interaction does not necessarily deteriorate the topological electromagnetic responses such as the anomalous Hall effect and the chiral magnetic effect in Weyl semimetals, for instance. 

The nonzero Chern number jump in the absence of band touchings exceeds the conventional picture of bulk-boundary correspondence, where the violation of adiabaticity is required to nullify the topological index. Only rarely does this happen in the noninteracting picture by reducing or enhancing the symmetry and the accompanied topological class\cite{gaplessTopo1}. More relevantly, this is caused by interaction effects\cite{gaplessTopo2}. Our case can be understood as topological numbers of interacting Green's functions that surpass the single particle picture. The zeros rather than the poles of the Green's function play the role of generating topological numbers\cite{topoGreenFunc1,*topoGreenFunc2}. In our case, this information is encapsulated in the complicated momentum dependence of the pairings in the self-consistent Hamiltonian, in contrast to constant gap-opening terms that cannot lead to the above topological feature. A complementary aspect of the robust Chern number jump is that exciton condensates prevent the gap closing\cite{FX1} which is necessary for a continuous phase transition between topologically trivial and nontrivial states.

\section{Concluding remarks}\label{Sec:Conclusion}
We study how the long-range Coulomb interaction affects the properties of a generic Weyl semimetal with the chemical potential at the Weyl nodes. 
There are recently some studies on the instability of Weyl semimetals with interaction\cite{WeylCDW1,Nandkishore,Aji2014,Wang2017}. In this paper, we provide a yet missing complete mean-field study considering all possible Coulomb interaction induced phases and let the self-consistent procedure manifest the major channel. 
The gap-opened phase has a charge density wave character from the viewpoint of translational symmetry breaking due to the internode interband coupling. The coherence of this coupling is s-wave like since the Coulomb interaction favors isotropic interband pairing.
Our main findings are that the Weyl nodes are not stable against strong enough Coulomb interactions while nontrivial topological Chern number jumps can survive after the gap opening. 
Our model itself cannot predict directly that anomalous Hall effect survives under strong Coulomb interaction, but it supports from a topological number viewpoint that topological responses are more robust than Weyl nodes themselves. A lattice model study of strong Coulomb interaction induced commensurate charge density wave order could explore whether a 3D magnetic insulator with nonzero Hall effect exists, which is beyond our current scope.

Some interesting questions for future studies might further include relating this topologically nontrivial state to the axionic predictions in a more direct manner and calculating the electromagnetic responses.
We also expect similar mean-field calculation could be done for the Dirac semimetal of much interest, using a doubled Hilbert space to account for the Kramers degeneracy. Besides the internode interband coupling dominant in this study, an intranode interband coupling is also possible to invalidate the symmetry protection and open the gap. The intranode coupling may not break translation symmetry but could break an $n$-fold rotational symmetry which leads to an interaction induced nematic state\cite{Juricic2017,FX3}.





\section*{acknowledgments}
The authors appreciate helpful discussions with A.H. MacDonald and M. Ezawa. X.-X.Z is grateful to Q. Niu for the hospitality during his stay at Austin (supported by the ALPS program).
This work was supported by ARO (No.~26-3508-81), the Welch Foundation (No.~F1473), JSPS Grant-in-Aids for Scientific Research (No.~26103006 and No.~16J07545) and CREST (No.~JPMJCR16F1).

\appendix
\section{Mean-field approxiamtion of the interaction}\label{App:MeanField}
Starting from the most general form of the Coulomb interaction 
$\hat{H}_\mathrm{I}=\frac{1}{2\Omega}\sum_{\vec{p},\vec{p}',\vec{q}}
V(\vec{q}) c_{\vec{p}+\vec{q}}^{\dagger} c_{\vec{p}'-\vec{q}}^{\dagger} c_{\vec{p}'} c_{\vec{p}}$, 
we expand the electron operator and make use of the reduced momentum $\vec{k}=\vec{p}-s\vec{K}$ to get $c_{\vec{p}} = c_{\vec{p}+\vec{K}}^L + c_{\vec{p}-\vec{K}}^R$ for the two nodes and find six terms allowed by the momentum conservation. In the spin basis, we further have $c_{\vec{k}}^s = \sum_{\sigma=\uparrow\downarrow} \ket{\beta_\sigma^s} c_{\sigma\vec{k}}^s$, then the resulting expression is given in a compact form by Eq.~\eqref{eq:interaction_Hamiltonian}. Performing the summation over node index $s'$ in Eq.~\eqref{eq:interaction_Hamiltonian}, we in fact obtain three terms
\begin{equation}
\begin{aligned}
&\hat{H}_\mathrm{I1}=
V(\vec{q}) c_{\sigma \vec{k}+\vec{q}}^{s\dagger} c_{\sigma' \vec{k}'-\vec{q}}^{s\dagger} c_{\sigma' \vec{k}'}^{s} c_{\sigma \vec{k}}^{s} \\
&\hat{H}_\mathrm{I2}=
V(\vec{q}) c_{\sigma \vec{k}+\vec{q}}^{s\dagger} c_{\sigma' \vec{k}'-\vec{q}}^{\bar{s}\dagger} c_{\sigma' \vec{k}'}^{\bar{s}} c_{\sigma \vec{k}}^{s} \\
&\hat{H}_\mathrm{I3}=
V(\vec{q}+2s\vec{K}) c_{\sigma \vec{k}+\vec{q}}^{s\dagger} c_{\sigma' \vec{k}'-\vec{q}}^{\bar{s}\dagger} c_{\sigma' \vec{k}'}^{s} c_{\sigma \vec{k}}^{\bar{s}},
\end{aligned}
\end{equation}
 wherein we neglect the common prefactor and summations $\frac{1}{2\Omega}\sum_{\vec{k},\vec{k}',\vec{q}}\sum_{\sigma\sigma', s}$ for simplicity. Firstly, by the Hartree approximation that contracts the direct operators, these three parts become 
\begin{equation}
\begin{aligned} 
&\hat{H}_\mathrm{Hatree1}= V(0) \rho_{\sigma'\sigma' \vec{k}'}^{ss} c_{\sigma \vec{k}}^{s\dag} c_{\sigma \vec{k}}^{s} \\
&\hat{H}_\mathrm{Hatree2}= V(0) \rho_{\sigma'\sigma' \vec{k}'}^{\bar{s}\bar{s}} c_{\sigma \vec{k}}^{s\dag} c_{\sigma \vec{k}}^{s} \\
&\hat{H}_\mathrm{Hatree3}= 2V(2\vec{K}) \rho_{\sigma'\sigma' \vec{k}'}^{s\bar{s}} c_{\sigma \vec{k}}^{s\dag} c_{\sigma \vec{k}}^{s},
\end{aligned}
\end{equation} 
respectively. Note that the first two Hartree terms cancel out because of the charge neutrality constraint which is reflected in the definition of the density matrices in Sec.~\ref{Sec:Model}. Secondly, by the Fock approximation that contracts the exchange operators, the three parts become 
\begin{equation}
\begin{aligned} 
&\hat{H}_\mathrm{Fock1}= -2V(\vec{k}-\vec{k}') \rho_{\sigma'\sigma \vec{k}'}^{ss} c_{\sigma' \vec{k}}^{s\dag} c_{\sigma \vec{k}}^{s} \\
&\hat{H}_\mathrm{Fock2}= -2V(\vec{k}-\vec{k}') \rho_{\sigma'\sigma \vec{k}'}^{\bar{s}s} c_{\sigma' \vec{k}}^{\bar{s}\dag} c_{\sigma \vec{k}}^{s} \\
&\hat{H}_\mathrm{Fock3}= -2V(\vec{k}-\vec{k}'+2s\vec{K}) \rho_{\sigma\sigma' \vec{k}'}^{\bar{s}\bar{s}} c_{\sigma \vec{k}}^{s\dag} c_{\sigma' \vec{k}}^{s},
\end{aligned}
\end{equation} 
respectively. Combining the above, we arrive at the mean-field Hamiltonian Eq.~\eqref{eq:MF}.

\section{Self-consistency equations}\label{App:SelfCon}
Here we write down the mean-field self-consistency equations of a general two-band model, which is accessible to analytic analysis. We assign a linearly dispersing noninteracting part to it.
\begin{equation}
H=\sum_{\vec{k}}(a_{c\vec{k}}^{\dagger}, a_{v\vec{k}}^\dagger)
(\xi_{\vec{k}} \sigma_z-\Delta_{\vec{k}} \sigma_x)
\begin{pmatrix} a_{c\vec{k}}\\a_{v\vec{k}}\end{pmatrix},
\end{equation}
where
\begin{equation}\label{eq:sc}
\begin{aligned}
&\xi_{\vec{k}}=v_Fk-\frac{1}{2\Omega}\sum_{\vec{k}'}V(\vec{k}-\vec{k}')(1-\frac{\xi_{\vec{k}'}}{E_{\vec{k}'}}),\\
&\Delta_{\vec{k}}=\frac{1}{2\Omega}\sum_{\vec{k}'}V(\vec{k}-\vec{k}')\frac{\Delta_{\vec{k}'}}{E_{\vec{k}'}},\\
&E_{\vec{k}}=\sqrt{\xi_{\vec{k}}^2+\Delta_{\vec{k}}^2}.
\end{aligned}
\end{equation}
Here we assume the chemical potential at the band touching point. We can rewrite Eq.~\eqref{eq:sc} in integral form and recognize that all terms depend on the magnitude of the momentum only:
\begin{equation}\label{eq:intsc}
\begin{aligned}
&\xi_{k}=v_Fk-\frac{e^2}{4\pi\epsilon}\int (1-\frac{\xi_{|\vec{k}-\vec{q}|}}{E_{|\vec{k}-\vec{q}|}})\sin\theta dqd\theta d\phi  ,\\
&\Delta_{k}=\frac{e^2}{4\pi\epsilon}\int \frac{\Delta_{|\vec{k}-\vec{q}|}}{E_{|\vec{k}-\vec{q}|}}\sin\theta dqd\theta d\phi   ,\\
&E_{k}=\sqrt{\xi_{k}^2+\Delta_{k}^2}.
\end{aligned}
\end{equation}
The $1/q^2$ in Coulomb interaction cancels with Jacobian factor $q^2$.
At $\vec{k}=0$, Eq.~\eqref{eq:intsc} can be simplified 
\begin{equation}
\begin{aligned}
&\xi_0=-\frac{e^2}{\epsilon}\int_{0}^{k_\mathrm{max}}(1-\frac{\xi_q}{E_q})dq\\
&\Delta_0=\frac{e^2}{\epsilon}\int_{0}^{k_\mathrm{max}}\frac{\Delta_{q}}{E_q}dq\\
&E_0=\sqrt{\xi_0^2+\Delta_0^2}.
\end{aligned}
\end{equation}
We immediately notice that if $\xi_{q},\Delta_{q},E_q \propto q$ (a possible self-consistent solution) at finite momentum, then zero-momentum Hamiltonian terms are all linearly proportional to the cutoff we choose. This reassures our discussion of the lack of intrinsic length scale in the main text.

%

\bibliography{reference.bib}  

\end{document}